\documentclass[aps,prl,twocolumn,a4paper,floatfix,superscriptaddress,showpacs,showkeys]{revtex4-1}

\usepackage[usenames,dvipsnames]{color}

\usepackage{dcolumn,amsmath,xspace}
\usepackage{graphicx,color}

\usepackage{epstopdf}

\newcommand{\six}{\ensuremath{6\!\times\!6 ~}}
\newcommand{\degC}{\ensuremath{^{\circ}\text{C }}}
\newcommand{\sixrt}{\ensuremath{(6\sqrt{3}\!\times\!6\sqrt{3})\text{R}30^\circ}~}
\newcommand{\rt}{\ensuremath{6\sqrt{3}}~}

\begin{document}
\title{Semiconducting graphene from highly ordered substrate interactions}

\author{M.S. Nevius}
\author{M. Conrad}
\author{F. Wang}
\affiliation{The Georgia Institute of Technology, Atlanta, Georgia 30332-0430, USA}
\author{A. Celis }
\affiliation{Laboratoire de Physique des Solides, UniversitŽ Paris-Sud, CNRS, UMR 8502, F-91405 Orsay Cedex, France}
\affiliation{Synchrotron SOLEIL, L'Orme des Merisiers, Saint-Aubin, 91192 Gif sur Yvette, France}
\author{M.N. Nair}
\affiliation{UR1 CNRS/Synchrotron SOLEIL, Saint-Aubin, 91192 Gif sur Yvette, France}
\author{A. Taleb-Ibrahimi}
\affiliation{UR1 CNRS/Synchrotron SOLEIL, Saint-Aubin, 91192 Gif sur Yvette, France}
\author{A. Tejeda}
\affiliation{Laboratoire de Physique des Solides, UniversitŽ Paris-Sud, CNRS, UMR 8502, F-91405 Orsay Cedex, France}
\affiliation{Synchrotron SOLEIL, L'Orme des Merisiers, Saint-Aubin, 91192 Gif sur Yvette, France}
\author{E.H. Conrad}\email[email: ]{edward.conrad@physics.gatech.edu}
\affiliation{The Georgia Institute of Technology, Atlanta, Georgia 30332-0430, USA}

\begin{abstract}
\textbf{While numerous methods have been proposed to produce semiconducting graphene, a significant bandgap has never been demonstrated.  The reason is that, regardless of the theoretical gap formation mechanism, disorder at the sub-nanometer scale prevents the required chiral symmetry breaking necessary to open a bandgap in graphene.  In this work, we show for the first time that a 2D semiconducting graphene film can be made by epitaxial growth.  Using improved growth methods, we show by direct band measurements that a bandgap greater than 0.5 eV can be produced in the first graphene layer grown on the SiC(0001) surface.  This work demonstrates that order, a property that remains lacking in other graphene systems, is key to producing electronically viable semiconducting graphene.}
\end{abstract}

\maketitle
\newpage
It is well known that the first graphene layer grown on the SiC(0001) surface is not electronic graphene.  That is, the first ``buffer'' graphene layer does not show the linear dispersing $\pi$-bands (Dirac cone) expected at the $K$-point of metallic graphene.\cite{Ohta_PRL_07,Emtsev_PRB_08,Riedl_PRL_09} The lack of $\pi$-bands in experimental band maps of the buffer layer\cite{Emtsev_PRB_08} supported the theoretical conclusion that sufficiently strong covalent bonds between the buffer layer and the SiC interface would push the graphene $\pi$-bands below the SiC valance band maximum.\cite{Varchon_PRL_07,Mattausch_PRL_07} Aside from these very early studies, research on the SiC graphene buffer layer faded and was subsequently eclipsed by a wide variety of other unsuccessful ideas to open a band gap in exfoliated or chemical vapor deposition (CVD)-grown graphene.\cite{Novoselov_Nat_12}  
\begin{figure}
\includegraphics[angle=0,width=8.5cm,clip]{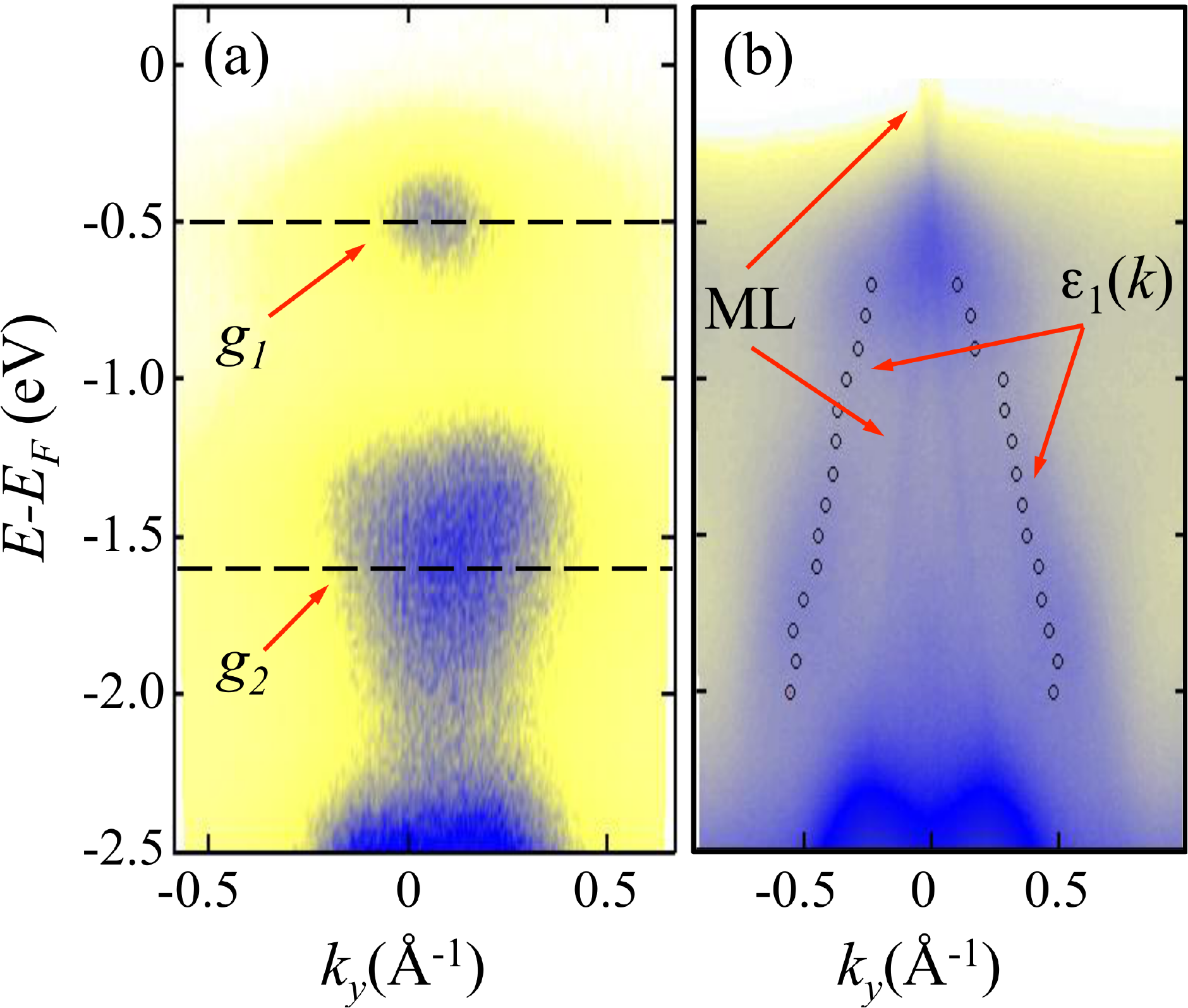}
\caption{(a) An ARPES cut through the graphene $K$-point of an under-grown \rt surface.  $k_y$ is perpendicular to $\Gamma\!-\!K$.  The states $g_1$ and $g_2$ (dashed lines) observed by Emtsev et al. [\onlinecite{Emtsev_PRB_08}] are marked.  (b) The same cut as in (a) for growth $20^\circ$C higher. Circles mark the peak positions along part of the $\epsilon_1$ band.}\label{f:Comp}
\end{figure}
\begin{figure}
\includegraphics[angle=0,width=7.5cm,clip]{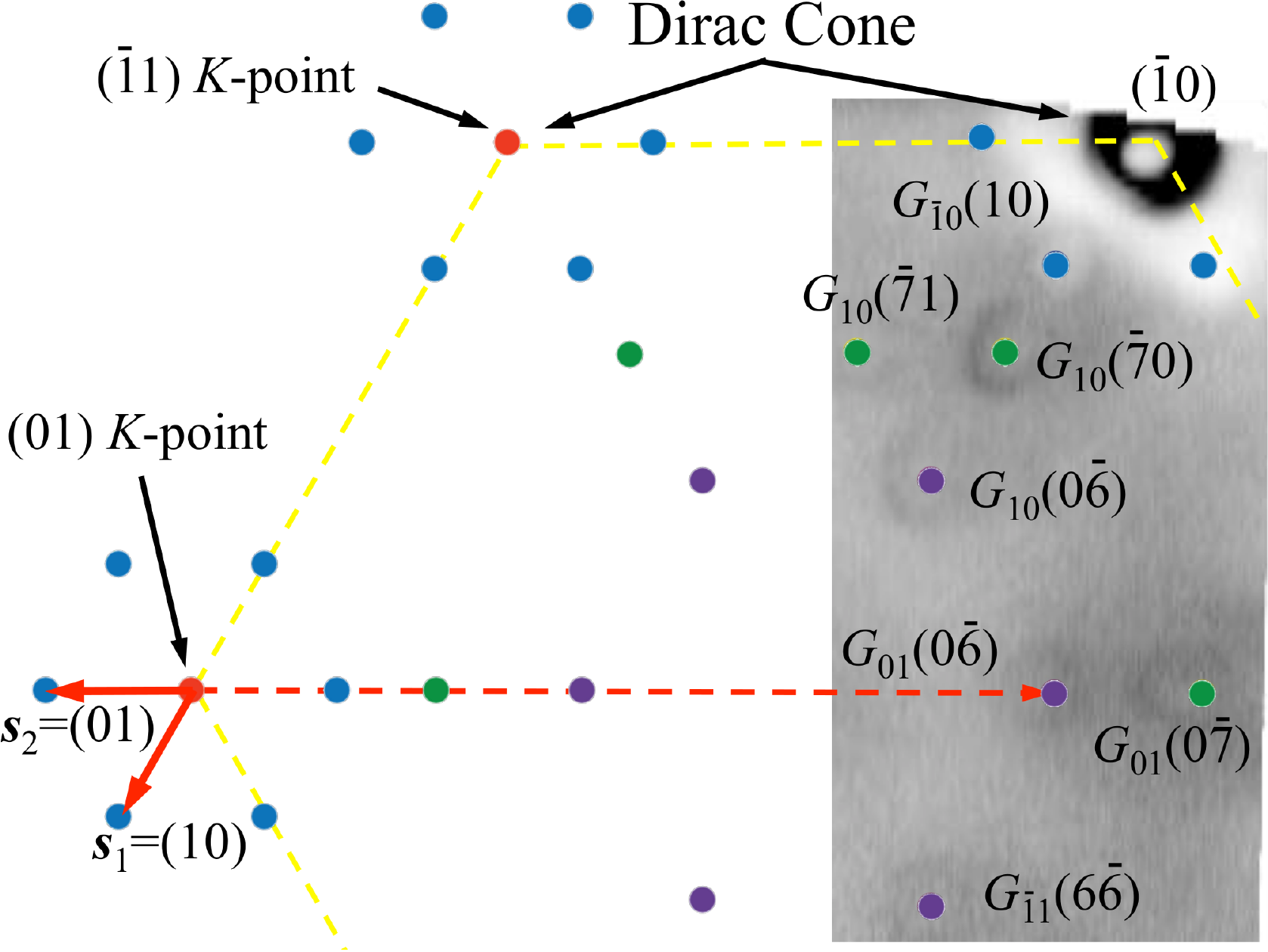}
\caption{A constant $E\!-\!E_F\!=\!-1$~eV cut through part of a ML graphene BZ showing replica cones. Blue dots mark  single \six umklapp scattering replicas of the Dirac cones $({\bf s}_1,{\bf s}_2)$.  Umklapp scattering of the Dirac cones from SiC $1\!\times\!1$ (purple dots) and SiC $1\!\times\!1$ plus \six reciprocal lattice vectors (green dots) are also shown.}
 \label{F:Replica}
\end{figure} 

\begin{figure*}
\includegraphics[angle=0,width=16.0cm,clip]{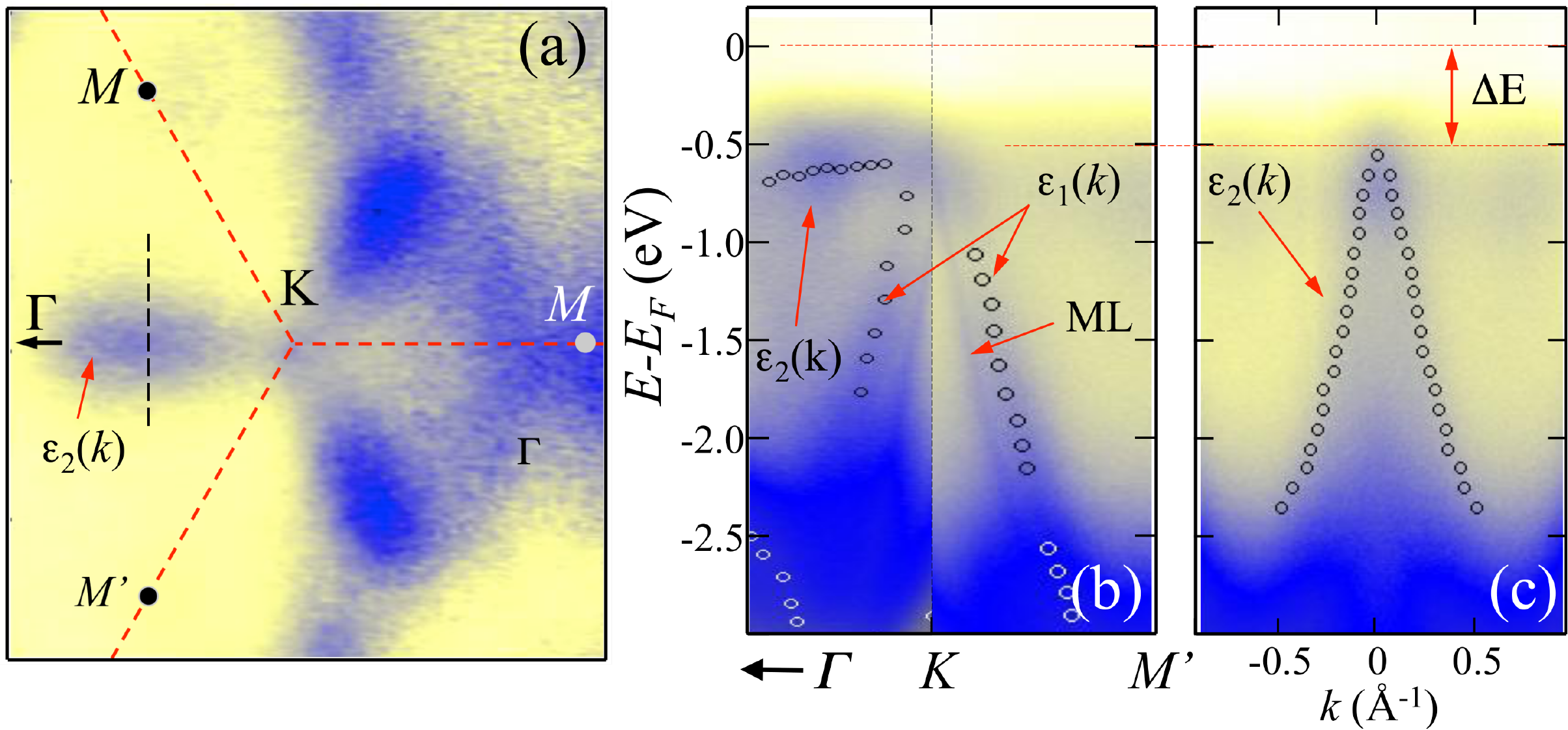}
\caption{(a) A constant energy cut through the graphene BZ near the K-point ($E-E_F=-0.41$eV, $h\nu=70$eV).  Dashed lines mark the boundary of the BZ. (b) A cut through the surface bands in the $\Gamma KM'$ direction. Circles mark the peak positions along part of the $\epsilon_1$ and $\epsilon_2$ band along with a few higher binding energy bands. A weak Dirac cone from a partial ML is shown. (c) (b) A cut perpendicular to $\Gamma K$ through the $\epsilon_2$ band [vertical black dashed line in (a)]. Circles mark the peak positions of the $\epsilon_2$ band.}\label{f:ARPES_FS}
\end{figure*}

One method to open a band gap in graphene is by periodic bonding to either all $A$ or all $B$ sites, which breaks graphene's chiral symmetry (referred to as graphene functionalization). The buffer graphene, commensurately bonded to the SiC(0001) surface, should have been an excellent example of a functionalized system that induces a bandgap. Despite the buffer graphene's potential to be functionalized by a commensurate and, most importantly, ordered array of Si or C atoms in the SiC, there was a major research shift to functionalize CVD-grown graphene.  Efforts to functionalize CVD graphene by a number of other methods have been a major research area.  As of this writing, no functionalized graphene, or graphene modified by any other proposed method, has been developed that produces a workable semiconducting form of graphene.  The problem with these methods is the inherent disorder introduced in the functionalization\cite{Bekyarova_JPhysD_12} and growth process.\cite{Novoselov_Nat_12} In fact, the lack of a graphene bandgap was the motivation to shift research to metal dichalcogenides despite the inability to grow them at the level of purity and order required for industrial scale electronics.

In this work, we use furnace-grown graphene to produce a structurally well ordered buffer graphene on the SiC(0001) surface. Angle resolved photoemission (ARPES) measurements show new dispersing $\pi$-bands that are not observed in samples grown by previous methods. These bands live above the SiC valance band maximum near the Fermi Energy, $E_F$.  The new band structure is a result of improved order caused by a higher growth temperature which, for the first time, gives rise to a well ordered $6\!\times\!6$ reconstruction in surface x-ray scattering experiments.\cite{Conrad_SXRD_tbp}  The bandgap, which is $>\!0.5$~eV, appears to be the result of the chiral symmetry breaking caused by the $6\!\times\!6$ reconstruction. The ARPES show that the buffer graphene layer on SiC is a true semiconductor, the goal of the first graphene electronics research.\cite{Berger04,Berger06}  

The substrates used in these studies were n-doped $n\!=\!2\times\!10^{18}\text{~cm}^{-2}$ CMP polished on-axis 4H-SiC(0001). The graphene was grown in a controlled silicon sublimation furnace.\cite{WaltPNAS} Graphene growth is a function of temperature, time, and crucible geometry that sets the silicon vapor pressure.  With the current crucible design, a monolayer (ML) graphene film will grow at 1520\degC in 20~min.  Using the same crucible, the semiconducting buffer layer discussed in this paper will grow at a temperature 160\degC lower than the ML in the same amount of time. Growing 20\degC lower than the optimum buffer temperature will still give the same \sixrt LEED pattern (referred to as \rt in the subsequent discussion) as the optimum buffer film but will not show the gapped $\pi$-bands discussed below.  

 Early ARPES work on the UHV-grown \rt reconstructed SiC(0001) surface (referred to as the graphene buffer layer in later literature) found that two non-dispersing states $g_1$ and $g_2$ at -0.5 and -1.6 eV were the only band features between $E_F$ and the SiC valance band maximum.\cite{Emtsev_PRB_08} These states were interpreted as localized Mott-Hubbard states hybridized from SiC surface dangling bonds.  We can reproduce these states by heating the SiC 20\degC cooler than the optimal buffer growth temperature.  Figure \ref{f:Comp}(a) shows an ARPES cut through the graphene $K$-point from this ``sub-buffer'' film.  The two surface states seen in previous work are clearly visible.  However, by heating 20\degC higher, a new dispersing band, $\epsilon_1(k)$ appears [see Fig.~\ref{f:Comp}(b)]. The new surface state is robust, being reproducible in multiple samples. Note that a faint linear Dirac cones appears at $k_y\!=\!0$.  This is due to a small amount of ML graphene  ($<\!2\%$) that typically nucleates at intrinsic step edges.\cite{Emtsev_Nat_09}  The Dirac point of the partial monolayer has the typical n-doping (0.55eV below $E_F$). 
 
Another indication of the improved sample order is the quality of the monolayer grown above the optimum buffer.  Figure \ref{F:Replica} shows a constant energy cut through part of the Brillouin Zone (BZ) of a ML graphene film.  In addition to the Dirac cone, replicas of the Dirac cone from umklapp scattering processes are also visible.  All replica cones from the $K^\text{th}$ $K$-point can be indexed using reciprocal lattice vectors of the SiC $6\!\times\!6$ unit cell: ${\bf G}_K(m,n)\!=\!m{\bf s}_1\!+\!n{\bf s}_2$, where  $|{\bf s}_1|\!=\!|{\bf s}_2|\!=\!\frac{1}{6}|{\bf{a}}^*_{SiC}|$ [see Fig.~\ref{F:Replica}].  In the ordered ML films, replica cones are clearly seen from both $1^\text{st}$-order in the \six unit cell $({\bf s}_1,{\bf s}_2)$ and from multiple scattering processes involving $1^\text{st}$-order $({\bf s}_1,{\bf s}_2)$ plus a SiC ${\bf G}$ vector (e.g., the ${\bf G}_K(\bar{7},0)$ and ${\bf G}_K(\bar{7},1)$). Early UHV grown samples only showed $1^\text{st}$-order replicas (i.e, $n, m\!=\!1$).\cite{Bostwick_NatPhys_07} The fact that so many ARPES replicas bands are observed in these films, along with the 6th order x-ray diffraction rods,\cite{Conrad_SXRD_tbp} testifies to the film's improved order.  

\begin{figure}
\includegraphics[angle=0,width=7cm,clip]{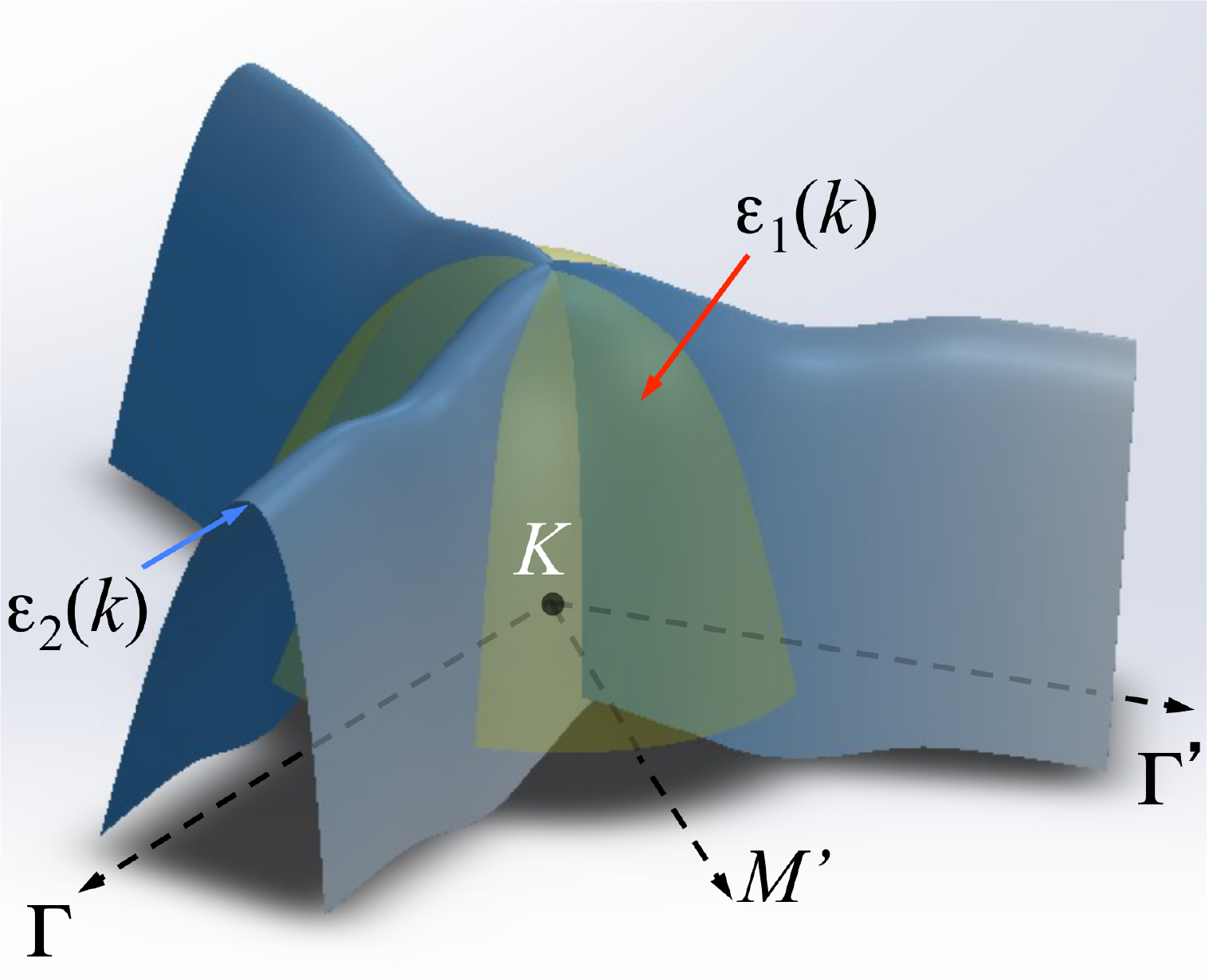}
\caption{
A schematic representation of the $\epsilon_1(k)$ and $\epsilon_2(k)$ buffer layer bands near the top of the $\pi$-bands around the $K$-point.}\label{f:Model}
\end{figure}

Detailed ARPES measurements from these improved samples show that the $\epsilon_1(k)$ band [Fig.~\ref{f:Comp}(b)] is a gapped graphene $\pi$-band.  Figure \ref{f:ARPES_FS}(a) shows a constant energy cut though part of the BZ of a buffer layer graphene near the $\epsilon_1(k)$ band maximum. Three lobes are visible. These lobes represent a second dispersing bands, $\epsilon_2(k)$, that is marked in the $\Gamma KM'$ cut in Fig.~\ref{f:ARPES_FS}(b) and (c).  Again, a Dirac cone from a small amount of ML graphene is visible.  The two bands are independent of the perpendicular momentum $k_\perp(E)$ and therefore cannot be due to bulk bands. The tops of both bands lie $\Delta E\sim\!0.5$~eV below $E_F$, or 1.8~eV above the valance band maximum of SiC interface, indicating that the buffer is a wide band gap semiconducting form of graphene.  A schematic of the two bands is shown in Fig.~\ref{f:Model}. The $\epsilon_1(k)$ band appears as a gapped $\pi$-band that disperses slower perpendicular to $\Gamma K$ than along either $\Gamma K$ or $KM$ directions [see Table \ref{tab:Band_Parm}].  The linear part of the $\epsilon_1(k)$ band has a velocity, $v$, that is significantly lower than the Fermi velocity, $v_F$, reducing to nearly half $v_F$ perpendicular to $\Gamma K$ [see Table \ref{tab:Band_Parm}].  
\begin{table}[htbp]
\caption{\label{tab:Band_Parm}The band velocity ($v$) and effective mass ($m^*$) near the $K$-point near the $\pi$-band maximum. $m^*$ is estimated assuming parabolic bands near $E_F$.}
\begin{ruledtabular}
\begin{tabular}{lccc}
  Band                          &$v/v_F$       &$m^*/m_e$ \\                          
 \hline 
ML  Dirac cone       				& 1.0                        &-\\
$\epsilon_1$ ($\perp \Gamma K$)     & $0.55\pm 0.01$       & $1.0\pm 0.02$\\
 $\epsilon_1$ ($\Gamma K$)     	 &  $0.63\pm 0.1$       & $1.5\pm 0.5$\\
 $\epsilon_1$ ($KM$)                    &  $0.80\pm 0.1$       & $0.55\pm 0.05$\\
$\epsilon_2$   ($\perp \Gamma K$)    & $0.98\pm 0.07$   & $0.25\pm 0.02$ \\
$\epsilon_2$     ($\Gamma K$)            & -                     & $1.5\pm 0.1$\\
 \end{tabular}
\end{ruledtabular}
\end{table}

The $\epsilon_2(k)$ band is 3-fold symmetric, extending towards $\Gamma$ and dispersing perpendicular to $\Gamma K$.  Figure \ref{f:ARPES_FS}(c) shows a cut perpendicular through the lobe in Fig.~\ref{f:ARPES_FS}(a).  The band velocity of $\epsilon_2(k)$ perpendicular to $\Gamma K$ is nearly the same as monolayer graphene [see Table \ref{tab:Band_Parm}].  The $\epsilon_1$ band has an effective mass ($m^*$) that ranges between 0.55 to 1.5$m_e$, while $\epsilon_2$ is a light band perpendicular to $\Gamma K$. 

In a broad sense, the gapped band structure near the $K$-point strongly suggests chiral symmetry breaking that mixes the $\pi$ bands from the $K$ and $K'$ points.\cite{McCann_PRL_06}  Any periodic potentials that break the $AB$ symmetry in the graphene through bond formation, chemical or strain fields, or finite size effects can open a gap in graphene. Weak interactions like those in bilayer graphene only produce small gaps.\cite{Ohta_Sic_06}  The strain  necessary to induce Kekule distortions,\cite{Okahara_ChemPLet} which produce band gaps of the order observed in these samples, are large enough to tear the graphene.\cite{Lee_Nano_11} That strain level is inconsistent with the 0.4\% strain measured by X-ray scattering.\cite{Schumann_PRB_14,Conrad_SXRD_tbp}  Strong bonding using Aryl functionalization\cite{Bekyarova_JPhysD_12,Bekyarova_JACS_09} has given large gaps, but the disorder inherent during the functional group's incorporation into the graphene lattice leads to poorly resolved band structure and low mobilities.\cite{Niyogi_NLet_10}  

A theoretical understanding of the buffer layer, and therefore an understanding of the origin of the observed gap, is difficult because of the excessive calculation time associated with exploring different models for the large \rt unit cell. Rather than calculating the full \rt cell, early calculations instead used an almost commensurate $\sqrt{3}\!\times\!\sqrt{3}$~R30 SiC reconstruction to make calculations more tractable.\cite{Varchon_PRL_07,Mattausch_PRL_07} These calculations predicted that strong $\text{sp}^3$ bonds between 2/3 of the interfacial Si atoms and the buffer buffer graphene caused the $\pi$-bands to shift above and below the conduction band minimum and valance band maximum, respectively.  The calculations also predicted a metallic, slightly delocalized, surface state near $E_F$ due to the remaining unbounded SiC Si atoms in the interface, similar to the states observed experimentally in the earlier, less ordered samples like in Fig.~\ref{f:Comp}(a).\cite{Emtsev_PRB_08}  These approximate models are clearly insufficient to explain the observed bands. Only one {\it ab inito} calculation by Kim et al.\cite{Kim_PRL_08} has calculated the band structure for the buffer using a full \rt cell. While the calculation was restricted to a bulk terminated SiC interface,\cite{Kim_PRL_08} it does give some insight into the origin of the observed gap when compared to the ARPES results. 
\begin{figure}
\includegraphics[angle=0,width=6.5cm,clip]{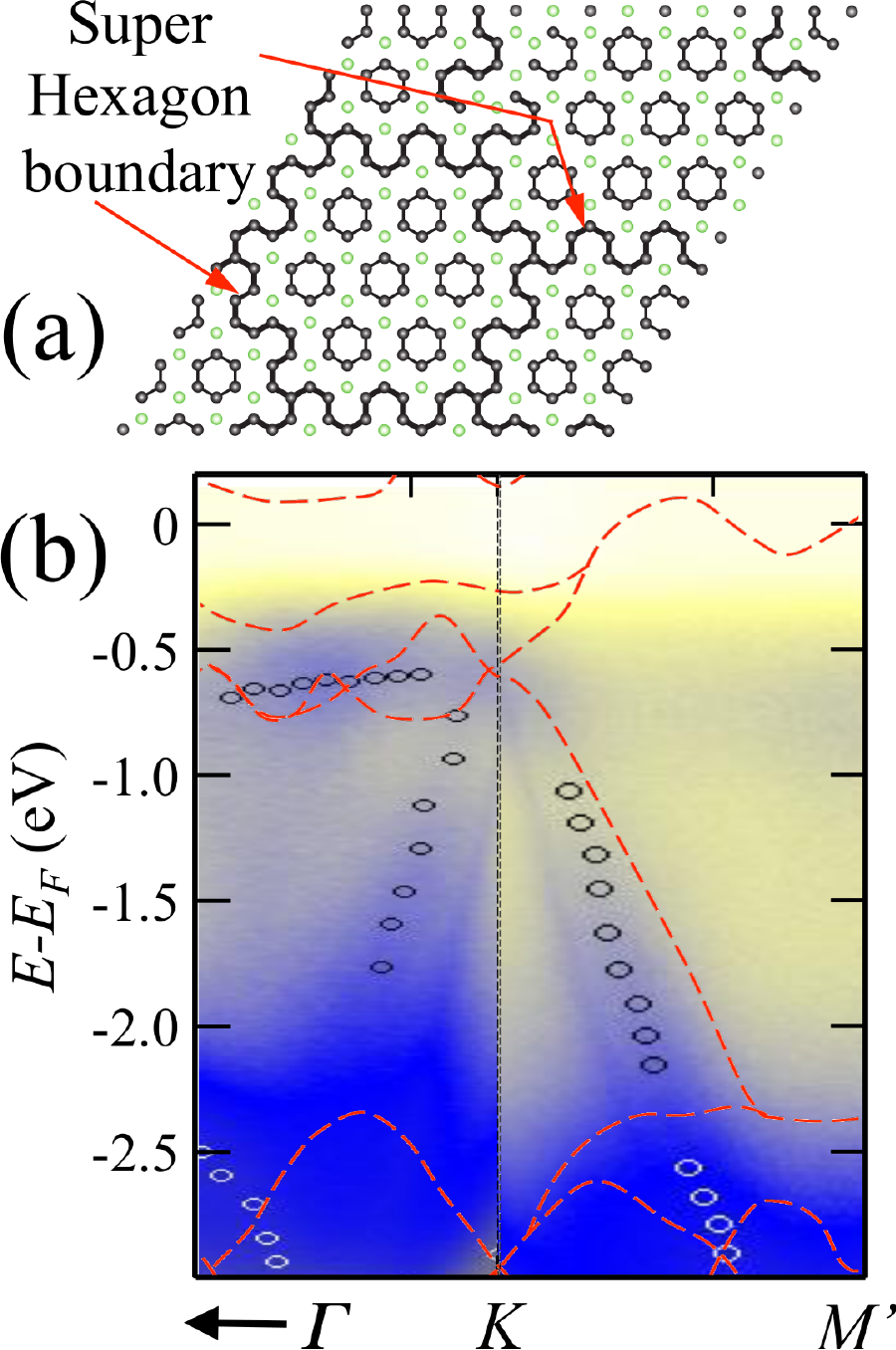}
\caption{(a) The calculated graphene buffer layer \sixrt cell on a relaxed bulk terminated SiC(0001) surface.\cite{Kim_PRL_08}  Green open circles are buffer carbon atoms that are bonded to the SiC surface.  Black circles are carbon not bonded to substrate.  Chains of atoms define superhexgon regions.  (b) An outline of the calculated bands (red dashed lines) from a bulk terminated SiC plus buffer layer from Ref.~[\onlinecite{Kim_PRL_08}] overlaid onto the experimental bands.  The theoretical bands have been shifted 0.13~eV lower to better match the $\epsilon_2$ band.}
 \label{F:Model}
\end{figure}

Kim et al.\cite{Kim_PRL_08} find that about 25\% of the carbon atoms in the buffer graphene are covalently bonded to Si atoms on the SiC interface.  The resulting structure is a hexagonal network of graphene ribbon-like structures with the remaining buffer carbon atoms covalently bonded to the SiC surface [see Fig.~\ref{F:Model}(a)].  Similar hexagonal networks, either structural or electronic, have been observed in Scanning Tunneling Microscopy or produced theoretically.\cite{Owman_SS_96,Riedl_PRB_07,Varchon_PRB_07} Kim et al.'s DFT calculations show that the $\pi$-orbitals of carbon atoms on the superhexagonal boundaries (or ribbons) give rise to several bands near the $K$-point above and below $E_F$. These bands are overlaid on our measured bands in Fig.~\ref{F:Model}(b). We have shifted the calculated bands by -0.13eV to match the $\epsilon_2(k)$ band maximum. Like the experimental $\epsilon_1(k)$ and $\epsilon_2(k)$ bands, the theoretical model shows that the covalent bonding to the SiC does not completely destroy the $\pi$-bands as earlier calculations predicted. Nonetheless, the calculations do not reproduce several important features of the experimental bands.  The calculations to not predict the formation of a band gap.  They also do not correctly reproduce the dispersion of the $\epsilon_1$ band, especially from $\Gamma$ to $K$.

We suggest that while the ribbon structure produced in the model of Kim et al.\cite{Kim_PRL_08} may be correct, the large amount of covalent bonds associated with a bulk terminated SiC likely over estimates the graphene-SiC interaction. It is more likely that the buffer graphene is bonded to the SiC through a smaller number of sites.  A lower number of graphene-Si bonds is more consistent with STM measurements that suggest the buffer lies above a small set of Si-trimers.\cite{Rutter_PRB_07}  A reduced buffer-SiC bonding geometry is also consistent with both x-ray scattering\cite{Hass_PRB_08} and x-ray standing wave experiments,\cite{Emery_PRL_13} which find a reduced Si-concentration and an increased C-concentration in the SiC layer below the buffer.  We suggest that reduced substrate bonding would still be sufficient to strain the buffer to produce the observed ribbon network.  The ribbon network, which is now isolated from the substrate, would result in a finite size induced band gap, as observed.\cite{Fujita_JPhy_Jap_96}  

In this work we show that, despite claims to the contrary, a semiconducting form of graphene can be manufactured.  Using improved growth methods, we have been able to produce a well ordered single layer of graphene on the SiC(0001) surface.  The first graphene layer, known as the buffer layer, is a semiconducting form of graphene.  Using ARPES, we show that the improved sample order leads to new bands with a band maximum that lies $\Delta E\!\sim\!0.5$~eV below the Fermi Level.  Depending on where the conduction band lies, the bandgap of this form of graphene must be $>\!0.5$~eV.  While no theoretical model predicts the measured bands, the experimental bands resemble those from a network of graphene ribbons that are distortions in the buffer layer.  The distortions would be due to strain strain relief in the film caused by a subset of carbon atoms in the buffer that locally bond to the SiC surface.


 \begin{acknowledgments}
This research was supported by the National Science Foundation under Grant No. DMR-1401193, DMR-1005880.  M. Nevious also acknowledges support from the NSF PREM program under grant DMR-0934142.  Additional support came from the Partner University Fund from the French Embassy.
\end{acknowledgments}

\end{document}